\documentclass[conference,composoc,9pt]{IEEEtran}

\usepackage[nocompress]{cite}
\usepackage[pdftex]{graphicx}
\usepackage{url}
\usepackage[bookmarks=true,breaklinks=true,colorlinks,linkcolor=black,citecolor=blue,urlcolor=black]{hyperref}
\usepackage[final]{microtype}
\usepackage{listings}
\usepackage{caption}
\usepackage{subcaption}
\usepackage{xcolor}
\usepackage{multibib}

\usepackage{lipsum}
\usepackage{courier}

\let\svthefootnote\thefootnote
\newcommand\freefootnote[1]{%
  \let\thefootnote\relax%
  \footnotetext{#1}%
  \let\thefootnote\svthefootnote%
}

\makeatother

\title{``It's a Trap!''---How Speculation Invariance Can Be Abused with Forward Speculative Interference}

\begin{document}

\author{\IEEEauthorblockN{Pavlos Aimoniotis}
\IEEEauthorblockA{\textit{Uppsala University} \\
pavlos.aimoniotis@it.uu.se}
\and
\IEEEauthorblockN{Christos Sakalis}
\IEEEauthorblockA{\textit{Uppsala University} \\
christos.sakalis@it.uu.se}
\and
\IEEEauthorblockN{Magnus~Sj\"alander}
\IEEEauthorblockA{\textit{Norwegian University of Science and Technology} \\
magnus.sjalander@ntnu.no}
\and
\IEEEauthorblockN{Stefanos~Kaxiras}
\IEEEauthorblockA{\textit{Uppsala University} \\
stefanos.kaxiras@it.uu.se}
}
 
\maketitle

\thispagestyle{plain}
\pagestyle{plain}

\begin{abstract}
Side-channel attacks based on speculative execution access sensitive data and use transmitters to leak such data during wrong-path execution. Speculative side-channel defenses have been proposed to prevent such information leakage. In one class of defenses, speculative instructions are considered unsafe and are \emph{delayed} until they become non-speculative. 

However, not all speculative instructions are unsafe: Recent work demonstrates that speculative invariant instructions are independent of a speculative control-flow path and are guaranteed to eventually execute and commit, regardless of the outcome of the performed speculation. Compile time information coupled with run-time mechanisms can then selectively lift defenses for \emph{Speculative Invariant} instructions, regaining some of the performance lost to “delay” defenses. 

Unfortunately, speculative invariance can be easily mishandled with \emph{Speculative Interference} to leak information using a new side-channel that we introduce in this paper. Recent work shows that younger \emph{speculative} instructions can \emph{interfere} with older non-speculative instructions that are bound to commit. This “backward” speculative interference reveals speculatively accessed secrets through the non-speculative instructions, in a way that delay-defenses do not cover, rendering them ineffective for this type of attack. 

In our work, we show that the counterpart to backward speculative interference, i.e., \emph{forward speculative interference}, enables \emph{older} speculative instructions to interfere with \emph{younger speculative-invariant (bound-to-commit) instructions}, effectively turning them into transmitters for secret data accessed during speculation. We demonstrate \emph{forward speculative interference} on real hardware, by selectively filling the reorder buffer (ROB) with spurious instructions, pushing speculative-invariant instructions in- or-out the ROB \emph{on demand}, based on a speculatively accessed secret. This reveals the speculatively accessed secret, as the occupancy of the ROB itself becomes a new speculative side-channel. We also demonstrate that it is possible to use the x86 ISA REP prefix, which unrolls as a micro-op loop in the microarchitecture at decode time (before any side-channel defenses have taken effect), as a method for generating spurious instructions. We propose several mitigations that range from changing compile-time decisions for speculative-invariance to run-time mechanisms that aim to make ROB occupancy operand-independent. 

\end{abstract}

\section{Introduction}

\freefootnote{A later version of this paper has been published on IEEE Computer Architecture Letters (CAL) under the name \textit{Reorder Buffer Contention}~\cite{aimoniotis2021reorder} and Digital Object Identifier no. 10.1109/LCA.2021.3123408.}

Speculative side-channel attacks use speculative execution to gain access to
information that would otherwise be inaccessible. %
Speculatively executed instructions are capable of temporarily bypassing
hardware or software defenses to gain illegal access to data that are then passed to speculative side-channel instructions,
a \emph{transmitter gadget}, capable of leaking those sensitive data to the non-speculative domain. %
Transmitter gadgets perform an operation that alters the
microarchitectural state of the processors, leading to a data leak.
A receiver observes the changes in the microarchitectural states and is
able to identify leaked data outside of the speculation window.

To tackle this problem several hardware defenses~\cite{dom19, yan_invisispec:MICRO2018, yu_speculative_2019, muontrap20, barber_specshield_2019, safespec_2019, Saileshwar19, taram_context-sensitive_2019}
have been proposed, introducing a variety of security guarantees. %
However, defenses also introduce various levels of complexity and performance overhead. %
Several hardware defenses rely on techniques that protect instructions while they are
speculative, and focus on making them invisible. %
One example is Delay-on-Miss (DoM)~\cite{dom19}. DoM delays speculative loads that miss in the
L1 cache until they become non-speculative, at which point they can be executed safely. %
Another example is InvisiSpec~\cite{yan_invisispec:MICRO2018}. %
InvisiSpec performs speculative loads but keeps the effects of a miss invisible in the cache hierarchy. %
When the speculation is verified, changes in the memory hierarchy are effected with a visible access.

Hardware defenses, such as DoM and InvisiSpec, add significant
performance overhead~\cite{dom19, yan_invisispec:MICRO2018}. %
For this, Zhao et al. proposed InvarSpec~\cite{invarspec20}, a framework that detects and lifts
the protection for speculative instructions that become \emph{speculation invariant}. %
For an instruction to be speculative invariant, its data and control
dependencies must be resolved during the speculation window. %
Such instructions are eventually going to execute with the same operands, even if
they are temporarily squashed due to misspeculation, and are, thus, considered safe to execute. %
Lifting the protection for speculation invariant instructions
enables the visible execution of an instruction while it is still under
speculation, maintaining the ``invisible speculative execution'' semantics of
defenses such as DoM or InvisiSpec while recovering significant performance lost
to these defenses. %

In a related development, Behnia et al. demonstrate that \emph{Speculative Interference}~\cite{specinterference21}
can break (under some assumptions) the DoM and InvisiSpec defenses. %
Up until now, the transmitter instructions were considered to be exclusively under speculative execution. %
With the introduction of Speculative Interference attacks, this has changed. %
In such an attack, the transmitter instructions are placed
before (in program order) the speculation window. %
Hence, the transmitter instructions can lie outside the protection of DoM or InvisiSpec defenses, as these are engaged only for instructions
that follow (in program order) the source-of-speculation instruction(s).
Since Speculative Interference is based on the fact that younger speculative instructions can influence the timing of older instructions,
it can consequently lead to information leakage even under speculative
defense mechanisms~\cite{specinterference21}. %

The key insight of our work is that speculation-invariant instructions are susceptible
to speculative interference from \emph{older} speculative
instructions: \emph{Forward Speculative Interference} (FSI).
To clearly differentiate between FSI and the speculative interference from \emph{younger} speculative instructions,
we refer to the latter as Backward Speculative Interference (BSI).
Using FSI, a new side-channel can be created by manipulating
the inclusion or exclusion of speculation-invariant instructions in the reorder buffer (ROB).
Other forms of forward interference are also possible and Behnia et al.~\cite{specinterference21} discuss how to delay instruction fetch with reservation station (RS) contention, called ${G^I}_{RS}$ in~\cite{specinterference21}. However, ${G^I}_{RS}$ concerns blocking of instruction fetch (and the front-end) which affects the I-Cache and is distinctly different from the \emph{ROB-contention} interference discussed here that concerns instruction execution.

We demonstrate FSI with \emph{ROB contention} on actual processors (Intel Sandy Bridge) and show how the ROB can be used as a side-channel.
Specifically, we show how, during speculation, we can selectively push in-or-out of the ROB load instructions that are on the---yet unknown---correct path of execution, leading to side-effects that remain observable after the speculation has been resolved.
These load instructions would be marked as speculative-invariant by InvarSpec, therefore the InvarSpec framework is susceptible to such a side-channel attack as well.

In addition to the attacks, we propose FSI ROB-contention mitigations from the speculative invariance point-of-view.
We propose two potential mitigations: conservatively considering at compile time instructions that are susceptible to ROB-contention interference as non-speculation-invariant and compile-time path balancing to prevent ROB-contention FSI. Finally, we briefly touch on making ROB contention, operant-invariant to manage the ROB side-channel in a more general approach.
Evaluation of our proposed mitigations is work in progress and we aim to report results in a future version of the paper.

\section{Background}

\subsection{Delay-on-Miss}
Delay-on-Miss (DoM) is a hardware defense mechanism against speculative
side-channel attacks, focusing on side-channels that abuse the memory
hierarchy~\cite{dom19}. %
Consecutively, side-channel attacks that do not focus on the memory hierarchy
are outside the scope of DoM and are not hindered by it. %

DoM operates on two fundamental principles. %
First, DoM delays transient loads until they become non-speculative. DoM introduces the concept of \emph{speculative shadows} to efficiently track the speculative state of instructions and discover the earliest time instructions become non-speculative, typically significantly earlier than reaching the commit stage (becoming head of the reorder buffer).

Second, DoM delays only loads that miss in the cache. Because reading data into a cache requires complicated interactions with the rest of the system, it is difficult to hide the side-effects of loads in the memory hierarchy on a cache miss, as demonstrated in prior solutions such as InvisiSpec~\cite{yan_invisispec:MICRO2018} and Ghost Loads~\cite{ghosts}. %
However, a cache hit requires only small modifications to the cache state (update of the replacement state etc.), which can be easily deferred for when the load is non-speculative. %
Thus, instead of delaying all loads, DoM allows loads that hit in L1 cache to execute under speculation,
while delaying any side-effects until the load becomes non-speculative. %

\subsection{Speculation Invariance: InvarSpec}

InvarSpec is not itself a speculative side-channel defense but rather a framework that detects when a speculative
instruction becomes \emph{speculation invariant} and upon detection lifts any existing
protections for the instruction~\cite{invarspec20}. %
InvarSpec consists of two main parts. The first part is a compiler technique that after
static analysis generates a \emph{safe set} (SS) for the instructions. The second part
is a hardware mechanism that at runtime designates an \emph{execution-safe point} (ESP) according
to the SS.

\begin{figure}[hbt]
    \begin{subfigure}[]{0.5\columnwidth}
       \begin{lstlisting}[captionpos=b,
                      label={lst:idg},
                      basicstyle=\footnotesize\ttfamily,
                      breaklines=true,
                      numbers=left,
                      numberstyle=\tiny\color{gray},
                      language=C++,
                      escapechar=|,
                      tabsize=2,
                      commentstyle=\color{gray},
                      keywordstyle=\color{blue},
                      xleftmargin=15pt]
if(cond){     // instr1
 si = load i; // instr2
}
a = load si;  // instr3
b = load j;   // instr4
	\end{lstlisting}
        \caption{Source Code}
    \end{subfigure}
    ~
    \begin{subfigure}[]{0.49\columnwidth}
        \centering
        \includegraphics[width=\textwidth]{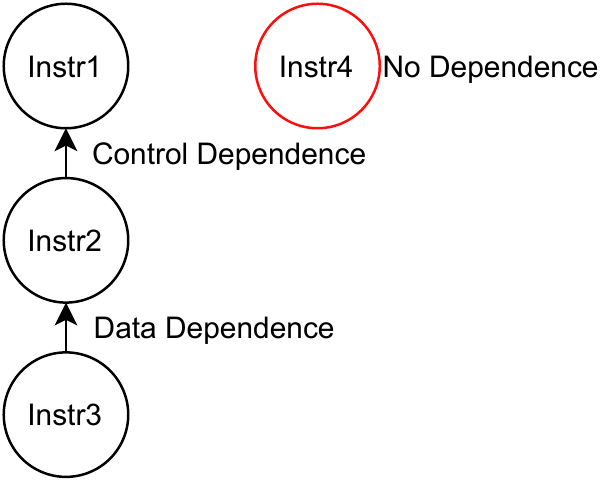}
        \caption{Instruction dependence graphs}
    \end{subfigure}

    \caption{Dependences related to safe set (SS)}
    \label{fig:invarspecidg}
\end{figure}

An example of speculation invariance is shown in \autoref{fig:invarspecidg},
where \texttt{a} (\textit{instr3}) has a potential data dependence
with \textit{instr2}, and \textit{instr2} has a control dependence with
\textit{instr1}. %
In order for \textit{instr3} to become speculation invariant, it must reach its
execution safe point, meaning both \textit{instr1} and \textit{instr2} must
reach their \emph{outcome safe point}. %
Since \textit{instr4} has no data nor control dependencies with any other
instruction (its SS is empty) it can execute immediately. %

Each instruction has its safe set (SS) defined by the compiler and corresponds
to the instruction's control and data dependencies on the instructions in the set~\cite{invarspec20}. %

The SS is used to determine at run-time when an instruction is ready and safe to execute
during speculative execution. %

An instruction is considered to be speculation invariant when it reaches its
execution-safe point (ESP). %
To reach the ESP, the operands of an instruction must have been finalized. %
Older instructions that comply with these rules are said to have reached their
outcome-safe point (OSP), meaning that their final result will not change, no
matter how many future squashes may happen. %
When everything in the safe set reaches the outcome-safe point, the instruction
itself has reached the execution-safe point and the speculative side-channel defense mechanisms can be lifted
for the instruction to be executed, even if the speculation has not be verified. %

\begin{figure}[hbt]
 \centering
\includegraphics[width=0.7\columnwidth]{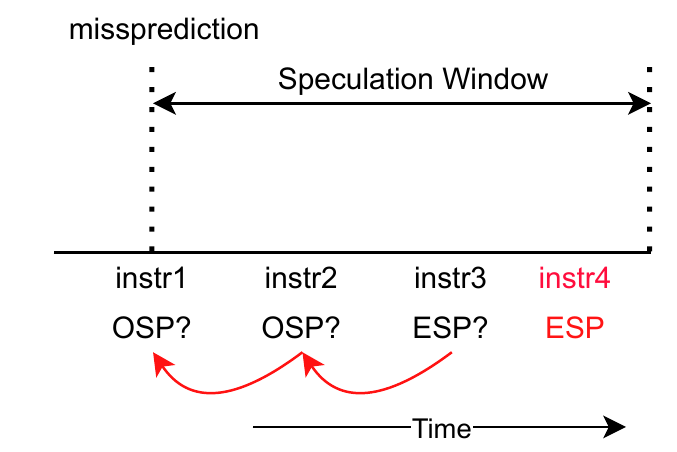}
    \caption{Speculation Invariant Timeline: For \textit{instr3} to be considered speculation invariant, \textit{instr2} and \textit{instr1} must reach their OSP. \textit{Instr4} has no dependences, and executes immediately under speculation using InvarSpec framework.}
    \label{fig:invarspectimeline}
\end{figure}

\autoref{fig:invarspectimeline} shows the timeline of an instruction using InvarSpec framework.
As a reminder, an instruction is said to have reached its ESP when all its operands reach their OSP. %
Once the instruction is ready to be executed, even if the speculation has not been resolved,
the defense mechanisms are lifted and the instruction executes. %

\subsection{Backward Speculative Interference}

Speculative Interference attacks~\cite{specinterference21} are able to break
defense mechanisms similar to DoM and InvisiSpec. %
Even though speculative loads are executed invisibly, misspeculated instructions
can change the timing of older instructions that \emph{may be outside the protection of DoM or InvisiSpec as non-speculative instructions}. %
This change can influence the ordering of memory operations that will be
committed, setting the fundamentals for a possible attack.

The attack consists of three parts: %
\begin{enumerate}
   \item A bound-to-commit instruction ---the \emph{interference target}--- waiting to be executed. %

   \item A branch predictor that is trained to mispredict, which creates a
      speculative window and the opportunity to illegally access some secret
      data. %

   \item The secret is used in an \emph{interference gadget} in such a way that
      the interference target is delayed in a secret-dependent manner. %
\end{enumerate}

For example, assume that the interference target is a load that takes $X$ cycles before its
operand becomes ready. %
The interference gadget can then use the secret value to selectively add contention in the MSHRs. %
For example, if the secret is equal to 1, the interference gadget attempts to fill
all MSHR entries before the interference target is ready to execute. %
Otherwise, if the secret is equal to 0, no memory operations are performed by
the interference gadget. %

Once the interference target becomes ready to execute, if the secret was 1 it will be further delayed,
otherwise, if the secret was 0, it will be executed unhindered. %
This difference in behavior can lead to information leakage as it can affect
the order of the interference target with respect to other loads, and thus
affect the cache replacement state. %

\section{ROB-contention: an FSI attack that breaks Speculative Invariance}
\label{section:forwardinterference}

Speculation invariance allows (bound-to-commit) speculative instructions to be executed without defenses
before the speculation is verified.
In  this respect, speculation-invariant instructions behave the same as the corresponding instructions in an unprotected processor.
In this work, we demonstrate our attack on an unprotected processor and then argue
that the same attack can be used to leak information on a processor that
implements InvarSpec.

In \emph{Backward Speculative Interference}, the interference gadget
delays the execution of the interference
target, a bound-to-commit instruction that is placed \emph{prior} to the speculation. %
In \emph{Forward Speculative Interference}, the interference gadget instead
interferes with a bound-to-commit speculation-invariant instruction, which
is executed \emph{while still under speculation}, unprotected by defense mechanisms like DoM~\cite{dom19} or InvisiSpec~\cite{yan_invisispec:MICRO2018}. %

\begin{figure}[hbt]

    \begin{subfigure}[]{0.23\textwidth}
       \begin{lstlisting}[captionpos=b,
                      label={lst:BSI},
                      basicstyle=\footnotesize\ttfamily,
                      breaklines=true,
                      numbers=left,
                      numberstyle=\tiny\color{gray},
                      language=C++,
                      escapechar=|,
                      tabsize=2,
                      commentstyle=\color{gray},
                      keywordstyle=\color{blue},
                      xleftmargin=15pt]
interference_target;
// mispredict
if(cond){
 interference_gadget;
}
	\end{lstlisting}
        \caption{Backward}
    \end{subfigure}
    ~
    \begin{subfigure}[]{0.23\textwidth}
       \begin{lstlisting}[captionpos=b,
                      label={lst:FSI},
                      basicstyle=\footnotesize\ttfamily,
                      breaklines=true,
                      numbers=left,
                      numberstyle=\tiny\color{gray},
                      language=C++,
                      escapechar=|,
                      tabsize=2,
                      commentstyle=\color{gray},
                      keywordstyle=\color{blue},
                      xleftmargin=15pt]
// mispredict
if(cond){
 interference_gadget;
}
interference_target;
	\end{lstlisting}
        \caption{Forward}
    \end{subfigure}

    \caption{Speculative Interference Attacks}
    \label{fig:spe-int-att}
\end{figure}

While FSI can take many forms, in this paper we introduce a novel side-channel based on manipulating ROB contention.
To the best of our knowledge, this has not been explored previously.
The ROB side-channel can be used to construct new Spectre~\cite{kocher_spectre_2018} variants on unprotected processors,
but more importantly, it can break InvarSpec approaches~\cite{invarspec20}
that selectively lift defenses of instructions under speculation.
Assuming DoM as the underlying defense mechanism---other defenses, such as InvisiSpec, are similarly susceptible---an FSI ROB-contention attack consists of three parts:
\begin{enumerate} %
\item A branch predictor that is trained to follow the attack path. %
\item A secret that is read from the cache (allowed in DoM) and ROB contention, as a function of the secret value, is added. %
\item A speculation-invariant target instruction that resides just after
the reconvergence point and that is executed with the DoM protections \emph{lifted}.
We initialize the speculation-invariant instruction with an empty safe
set, i.e., a set that has no dependencies and can execute immediately
when it becomes ready. %
\end{enumerate}
Depending on the contention-induced delay, and thus on the secret value, the speculation
invariant target instruction will be affected in terms of \emph{when it will be
ready to execute}. %
For example, when the secret is equal to 1, we add extra ROB contention,
in the form of a loop or a long sequence of spurious instructions. %
As a result, the ROB is filled with speculative instructions, which
prevents the speculation-invariant target instruction from even entering the ROB
and executing. %
On the other hand, the path followed when the secret is 0 behaves normally, enabling the speculation-invariant target
instruction to execute when it enters the ROB. %
Since InvarSpec has lifted the defenses from the instruction, any side-effects
caused by its execution will remain observable even after the
misspeculation has been detected and squashed, making it possible to infer the
secret value outside of the speculative window.

While the FSI ROB-contention attack shares some similarities with the ${G^I}_{RS}$ speculative interference attack, described by Behnia et al.~\cite{specinterference21}, it is distinctly different in a number of ways: First, in contrast to ${G^I}_{RS}$, ROB-contention manipulates the execution of bound-to-commit loads (which lie after the reconvergence point) rather than instruction fetch. As such, ROB-contention directly affects mitigations such as DoM or InvisiSpec (when combined with InvarSpec) that aim to protect \emph{data caches} from leaking information, which is not a concern with ${G^I}_{RS}$:
${G^I}_{RS}$ uses the instruction cache as a side-channel---ROB-contention uses the data cache. Second, ${G^I}_{RS}$ must cause a front-end stall to work. ROB-contention works as long as a target instruction is kept just outside the ROB, which does not necessarily mean a front-end stall. For example, if the target instruction is sufficiently far from the reconvergence point, the front end will keep fetching and decoding instructions from the reconvergence point onwards.

\begin{figure}[hbt]

   \begin{subfigure}[]{0.49\columnwidth}
        \centering
        \includegraphics[width=\columnwidth]{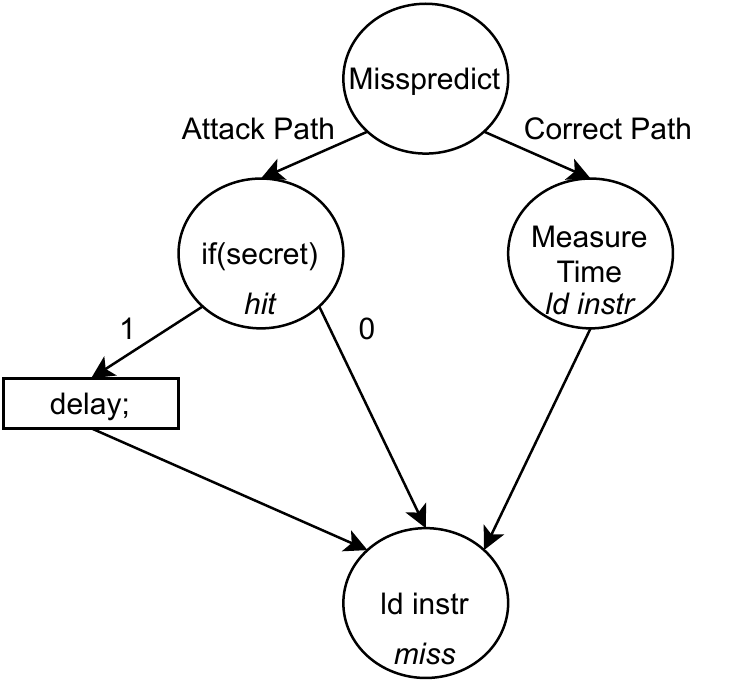}
        \caption{FSI v1: 
          Depending on the secret,
          we influence the execution time of the speculation-invariant
          target instruction, hence the latency of the measured instruction.}
    \label{fig:attackmeastime}
    \end{subfigure}
    ~
    \begin{subfigure}[]{0.49\columnwidth}
        \centering
        \includegraphics[width=\columnwidth]{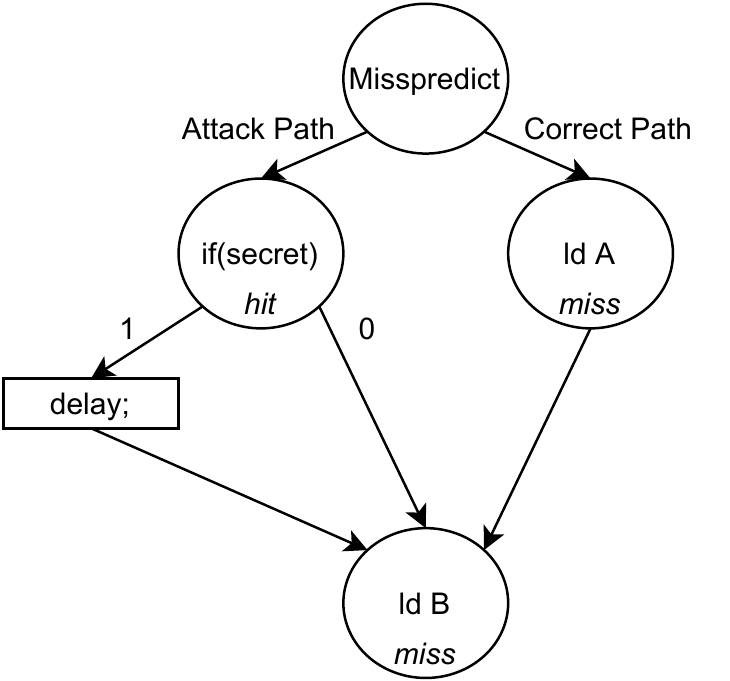}
        \caption{FSI v2: 
          Depending on the secret, %
          \textit{Ld A} and \textit{Ld B} will be placed either as \textit{Ld B, Ld A} %
          or \textit{Ld A, Ld B} in the cache.}
    \label{fig:attackcachepos}
    \end{subfigure}

    \caption{Two techniques to extract the secret.}
    \label{fig:attackgraphs}
\end{figure}

Two possible techniques to identify the secret, are shown
in \autoref{fig:attackgraphs}. %
The first technique can be thought-of as a version of the \emph{Flush\&Reload} attack~\cite{yarom_flush+_2014}.
It is shown in \autoref{fig:attackmeastime} and is
based on testing if data are cached in
the L1 cache or not. %
To achieve this, we measure the access time of the
speculation-invariant target instruction when the speculation is
finally resolved and the execution continues from the correct path. %
While on the misspeculated attack path, whether the load instruction at the
reconvergence point will be executed depends on which path the speculative
execution followed, i.e., it depends on if secret is 0 or 1.
Then, on the correct path, the time it takes to execute the load will change
depending on if the data was loaded by the attack path, thus making it possible
to infer the secret value.
The second technique (\autoref{fig:attackcachepos}), taken from Behnia et
al.~\cite{specinterference21}, is similar to the first technique but is instead
based on the relative order of two load instructions, as seen by the cache,
which causes visible changes in the cache replacement state. %
To do so, we load another address in the correct path that conflicts
with the address loaded by the speculation-invariant instruction. At a later time, we observe the cache
replacement state to extract the leaked information. %
We will discuss both of these techniques, as well as an alternative method (to
loops) for introducing ROB contention in the sections that follow.

\subsection{Measuring Cache Access Time}
\label{measuretime}
In this technique (\autoref{fig:attackmeastime}), we measure the access time of the
speculation-invariant target instruction when we access it during the
correct path, once the speculation is verified as incorrect. %

The attack starts by ensuring that the address of the
speculation-invariant target instruction is flushed from the cache. %
If the secret is equal to 1 then the speculative-invariant target
instruction is never executed along the incorrect path. %
Once the speculation is resolved and the correct path is taken a load
with the same address as the speculative-invariant target instruction
will miss in the cache and experience a long delay. %
If the secret is equal to 0 then the speculative-invariant target
instruction is executed in the incorrect path and the load in the
correct path will hit in the cache and experience a short delay. %

\subsection{Observing Replacement State}

In this technique (\autoref{fig:attackcachepos}), we focus on the replacement state of the L1 cache
to leak the information. %

Behnia, et al. demonstrate a complete attack based on replacement state~\cite{specinterference21}, but here, for simplicity, we assume a direct-mapped cache and load addresses that map to the same set.\footnote{Changes in a cache set or in its replacement state can be easily verified in a simulated environment, e.g., using gem5.}

Just as with the previous technique (\autoref{measuretime}), depending on the secret our goal
is to add ROB contention to affect the speculation-invariant instruction. %
Once the branch is resolved as incorrect and continues on the correct
path, another load instruction accesses a different address
that conflicts with the speculation-invariant load instruction on the
same cache set. %

If the secret is equal to 1, the speculation-invariant load
will be prevented from executing due to ROB contention. %
The load instruction from the correct path will be executed first,
taking place in the cache set. %
After that, the interference target (speculation-invariant load), will be normally
executed, evicting the load of the normal path from the cache. %
On the other hand, if the secret is equal to 0 then the
speculation-invariant load will be executed first, and
once the branch is resolved as incorrect,
the load from the correct path will evict the interference
target (speculation-invariant load) from the cache. %

\subsection{ROB Attack Using REP Instructions}
\label{section:robattack}

An FSI ROB-contention attack requires filling the ROB with speculative instructions.
While either a tight loop, or a long sequence of spurious instructions, fit the bill for this purpose,
interestingly, one can achieve the same result with a \emph{single static instruction}.

In the x86 ISA, \texttt{REP} is a prefix that can be used before string instructions. %
It creates a single-instruction loop,
with the value stored in the \texttt{ECX} register acting as the loop counter. %

The key property that enables a single \texttt{REP} instruction to affect ROB contention is that it unrolls as a $\mu$op loop in
the microarchitecture, at \emph{decode time}~\cite{angerfog21-instructiontables}.
ROB occupancy becomes a function of \texttt{ECX}. %

According to empirical studies~\cite{angerfog21-instructiontables,angerfog21}, \texttt{REP}-prefixed x86 instructions expand into a number of
$\mu$ops in the ROB.
The following table lists the $\mu$op expansion (\texttt{ECX==}\textit{n}) in the ROB for two typical \texttt{REP} instructions and for some well-known microarchitectures---similar expansion takes place for the majority of x86 microarchitectures~\cite{angerfog21-instructiontables}.

\begin{center}
\footnotesize
\begin{tabular}{ |c|c|c|c|c|c| } 
\hline
Instr./Proc.  & Haswell & Broadwell & Skylake & IceLake \\
\hline
\texttt{rep movs} & \textit{2n} & \textit{2n} & \textit{2n} & \textit{2n} \\ 
\hline
\texttt{rep lods} & \textit{5n+12} & \textit{5n+12} & \textit{5n+12} & \textit{5n+12} \\ 
\hline
\end{tabular}
\end{center}
\vspace{0.1cm}

Furthermore, we ascertain that the \texttt{REP movs} instruction expands \emph{speculatively} on a Sandy Bridge
microarchitecture.
We tested this scenario by giving \texttt{ECX} various values, after a speculation point, followed by a \texttt{REP} instruction (as in the code shown in~\autoref{fig:rep-code}).
By timing the code, we observe that the \texttt{REP} instruction, indeed, expands speculatively
into a number of $\mu$ops that is proportional to \texttt{ECX}. 

To mount a ROB attack with \texttt{REP} instructions (\autoref{fig:rep-code}), 
we use the speculatively-accessed secret to update the \texttt{ECX} register,
which then controls the number of $\mu$ops that are dispatched to the ROB. %
To create a large enough repetition factor, we left-shift the
secret by, e.g., ten places (if the secret is zero, it does not change). 
This value is passed to \texttt{ECX} which subsequently drives a \texttt{REP movs}
instruction to \emph{selectively} flood the ROB with up to $2n$ $\mu$ops. %

\begin{figure}[hbt]
 \centering
       \begin{lstlisting}[captionpos=b,
                      label={lst:rep},
                      basicstyle=\footnotesize\ttfamily,
                      breaklines=true,
                      numbers=left,
                      numberstyle=\tiny\color{gray},
                      language=C++,
                      escapechar=|,
                      tabsize=2,
                      commentstyle=\color{gray},
                      keywordstyle=\color{blue},
                      xleftmargin=15pt]
if(value){ // mispredict - Attack Path

 secret = secret << 10; // Repetition factor
 
 // Pass secret to ECX and execute rep
 asm("movl %0, %%ecx" : : "c" (secret));
 asm("rep movsb"); 
}
else { // Normal Path
 t1 = __rdtscp(); // Start measuring latency
 transmitter = probe[0]; // Evaluation
 t2 = __rdtscp(); // End measuring latency
 t = t2-t1;
}

transmitter = probe[0]; // Recovergence Point
	\end{lstlisting}
\caption{Abusing InvarSpec with Forward Speculative Interference using REP Instruction}
\label{fig:rep-code}
\end{figure}

\section{Attack Demo and Experimental Results}

We implemented our FSI attack on actual hardware. While DoM defenses and InvarSpec are not implemented, 
we can see the effects of the attack in an unprotected core, which behaves the same as a protected core with respect to speculative-invariant instructions.
We evaluated our results on an Intel\textsuperscript{\textregistered{}} Core\textsuperscript{TM} i7-2600K, %
which is a Sandy Bridge microarchitecture, running at up to 3.40GHz. %
The processor has 4 cores (2 SMT threads per core, for 8 threads in
total) and 3 cache levels. %
Each core has a 32KiB L1 Cache and a 256KiB L2 Cache, and all cores share an  8MiB LLC. %
Our source code is written in C, and we measure the timing of a variable
assignment to detect difference in the correct path as shown below. %
       \begin{lstlisting}[captionpos=b,
                      label={lst:eval},
                      basicstyle=\footnotesize\ttfamily,
                      breaklines=true,
                      numbers=left,
                      numberstyle=\tiny\color{gray},
                      language=C++,
                      escapechar=|,
                      tabsize=2,
                      commentstyle=\color{gray},
                      keywordstyle=\color{blue},
                      xleftmargin=15pt]
t1 = __rdtscp(); // Start measuring latency
transmitter = probe[0]; // Evaluation
t2 = __rdtscp(); // End measuring latency
t = t2-t1;
\end{lstlisting}

\begin{figure}[hbt]
\centering
    \includegraphics[width=\columnwidth]{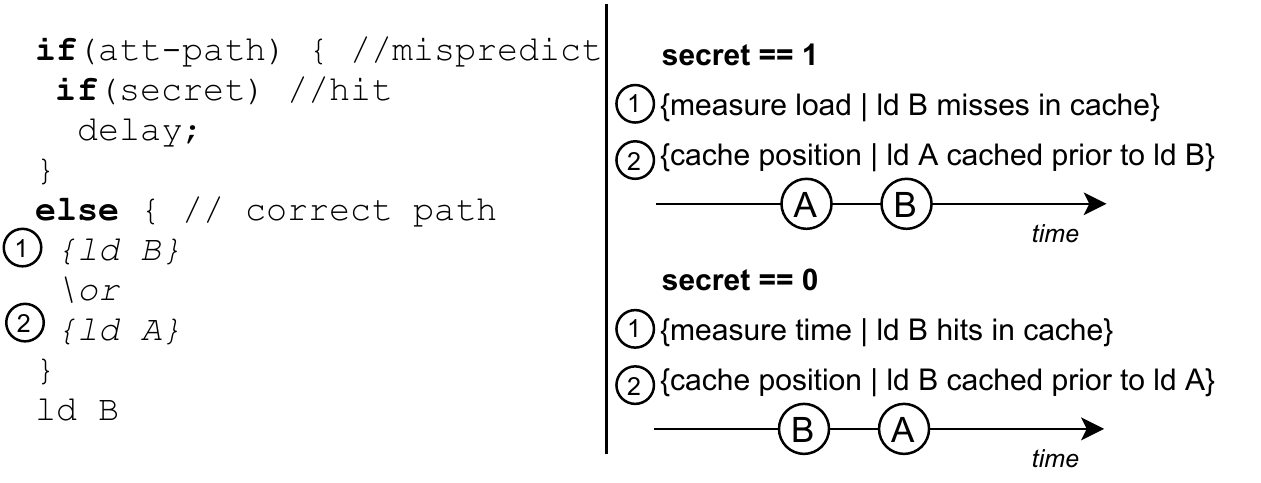}
    \caption{Attack Demo: %
    (1) measure time \textit{Ld B}, to see if its in cache and distinguish 
    the secret. (2) determine in Cache if A or B is cached. i.e. in a
    direct mapped cache where A and B maps on the same set, if \texttt{secret==1}, 
    B will evict A from the cache.}
    \label{fig:forwardSIatt}
\end{figure}

The overall structure of the attack demo is illustrated in \autoref{fig:forwardSIatt} for two variants: timing loads and determining the order of loads. %
We report on the results for the timing-load variant on a real
system. %
While determining the order of loads can be easily demonstrated in gem5,
on actual systems, this requires detection code (as in Behnia et al.~\cite{specinterference21}),
which is work in progress. %
Before we follow the attack path, all load addresses are flushed from %
the cache. %
The branch predictor is trained so that it will always mispredict and follow
the attack path. 
The secret value is already cached in the L1. 
Depending on the secret, ROB contention is added, so that \texttt{ld B} will be
delayed. %
If \texttt{secret==1}, delay from ROB contention will be sufficient for speculation to be verified
before \texttt{ld B} executes. %
If \texttt{secret==0}, no delay is applied and \texttt{ld B} 
is executed as soon as possible. %

\begin{figure}[hbt]
    \centering
    \includegraphics[width=0.6\columnwidth]{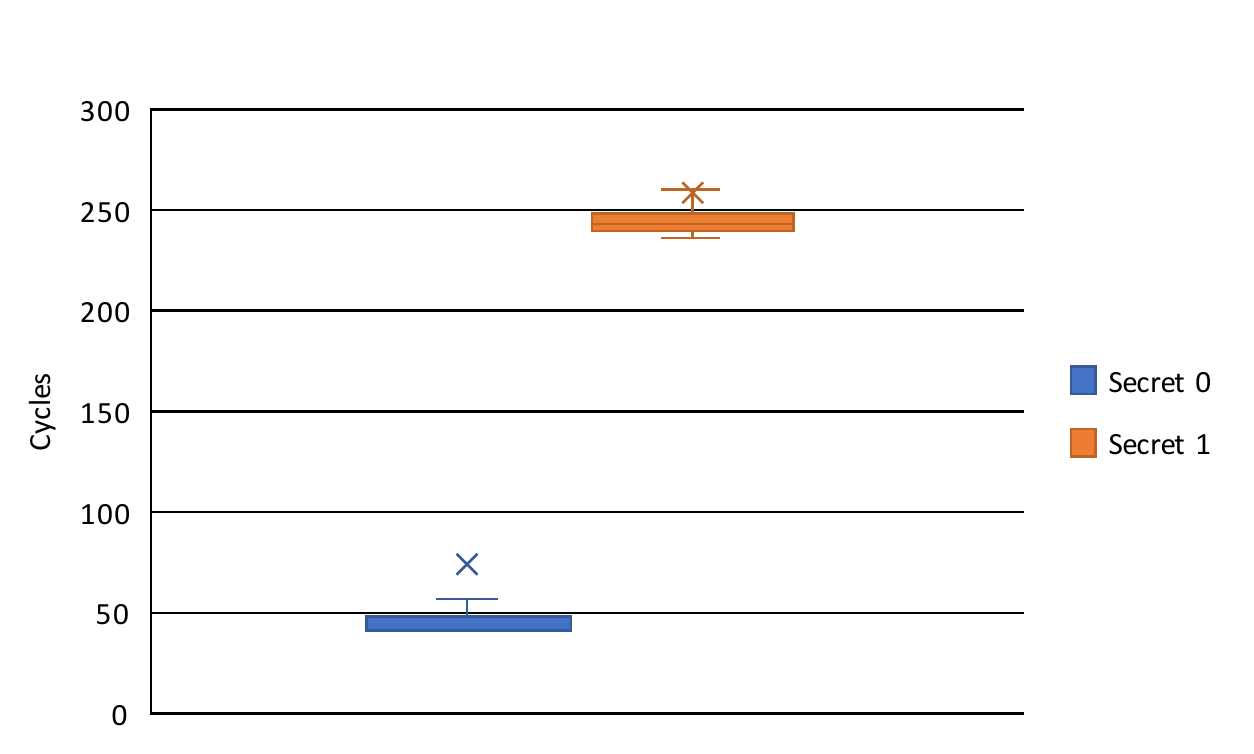}
    \caption{Attack using for-loop to delay the reconvergence point.}
    \label{fig:expresultsite}
\end{figure}

\subsection{Delaying with a for-loop}

In this version, we use a for-loop to delay the execution of the reconvergence point. %
This is the attack discussed in \autoref{section:forwardinterference}. %
The results are shown in \autoref{fig:expresultsite}. %
The distribution of all 1000 attempts per secret is presented, without outliers, in the graph. %
We show that an average load is equal to 258 cycles, if the secret is equal to 1,
and 73 cycles when the secret is equal to 0. %

\begin{figure}[hbt]
           \centering
\includegraphics[width=0.6\columnwidth]{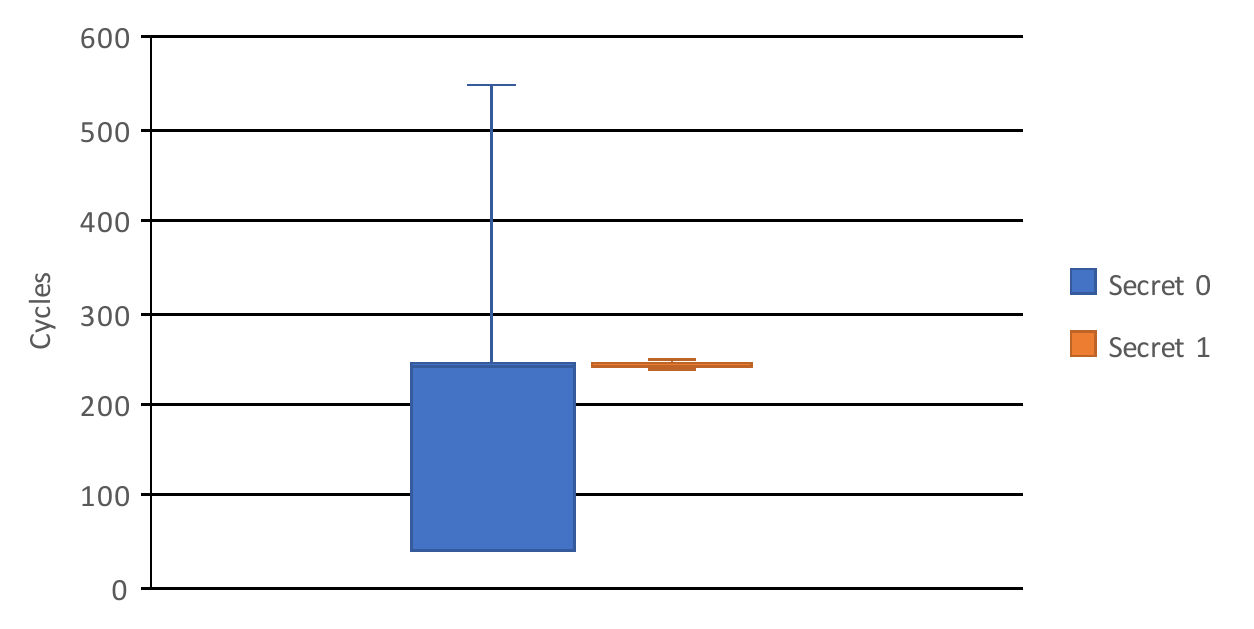}
    \caption{ROB Attack using REP instruction: All 1000 attempts per secret}
    \label{fig:Arobattack}
\end{figure}

\begin{figure}[hbt]   
           \centering
\includegraphics[width=0.6\columnwidth]{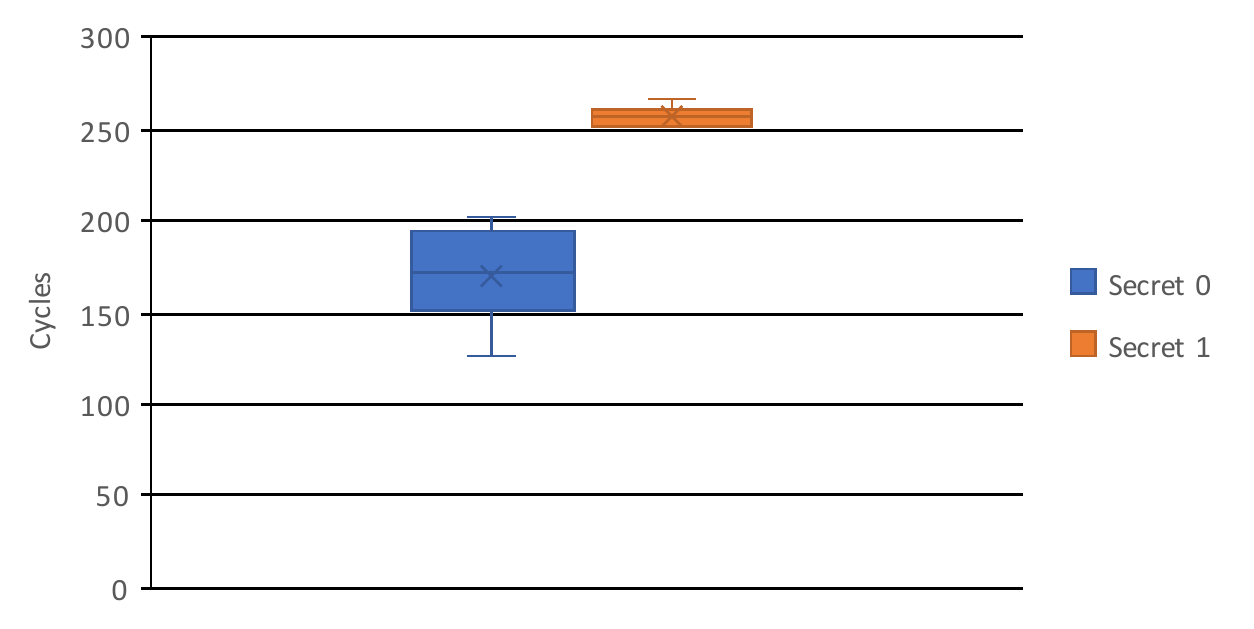}
    \caption{ROB Attack using REP instruction: Average every 100 attempts}
    \label{fig:Brobattack}
\end{figure}

\subsection{Delaying with REP instruction}

In this version, we use a \texttt{REP} instruction to delay the execution of the
reconvergence point. %
This is the attack discussed in \autoref{section:robattack}. %
\autoref{fig:Brobattack} illustrates the average of every 100 attempts. %
We show, that when repeating the attack, the results diverge, making it easier
to identify the secret: %
An average load when \texttt{secret==0} is 170 cycles. %
On the other hand, when \texttt{secret==1} an average load is 260 cycles. %

\subsection{Discussion}

Our results show that \emph{forward speculative interference} and ROB-contention work successfully in actual processors,
and constitute a new side-channel that can be used to construct Spectre-type attacks.
Because the speculation-invariant instructions behave the same as instructions from the re-convergence path in unprotected processors,
FSI ROB-contention poses a significant threat when we want to lift defenses for speculation-invariant instructions.

\section{Forward Speculative Interference Mitigations}

In this section, we discuss some possible mitigations for the FSI ROB-contention side-channel. %
Evaluation of these mitigations is currently work-in-progress and results will be presented in future versions of this paper.
We propose mitigations that are specific to InvarSpec+DoM, but also more general mitigations that can be applied to unprotected processors.

\paragraph{InvarSpec+DoM Specific Mitigations}
To protect against FSI ROB-contention attacks, InvarSpec must be conservative in declaring instructions as speculation-invariant if they are vulnerable to ROB-contention. For example, if the compiler can ascertain that the paths to the reconvergence point can differ in length,
it would be prudent \emph{not to} call instructions after the reconvergence path as speculation-invariant and take them off DoM protection.
This would include the cases described in our work: i.e., detecting loops or \texttt{REP} instructions in a path would automatically make that path suspect for ROB-contention as a variable-length path. Note that this mitigation can only reduce the benefit from InvarSpec but cannot introduce any overhead to a baseline DoM-protected system.
However, because in most cases paths to reconvergence do differ in length, this may lead to a dramatic reduction in coverage, i.e., the number of instructions that could be safely called speculation-invariant. One way to improve coverage is for the compiler to try to ``balance'' path-length differences (especially if these are modest) with padding, which may introduce small overheads in the shorter paths. We shall examine such options in future work.

\paragraph{General Mitigations}
Another direction to defend against FSI ROB-contention, independently of InvarSpec and delay defenses, is to make speculative ROB-filling \emph{operand-independent}~\cite{sdo20}. In other words, to ensure that the number of $\mu$ops that enter the ROB cannot be dependent on speculatively-accessed values. Similar to SDO~\cite{sdo20}, ROB-filling itself can be turned into a prediction, which can lead to a squash if mispredicted.

\section{Conclusion}

In this work, we present a new side-channel, based on ROB contention, and a new speculative execution attack (ROB-contention attack)
using this side-channel. The attack is achieved through \emph{Forward Speculative Interference}, i.e., 
speculative instructions interfering with \emph{younger} instructions that are bound to commit regardless of the speculation outcome. %
For this reason, techniques, such as the InvarSpec framework, that lift the defenses for such bound-to-commit instructions, are susceptible to the same attack and can leak speculatively accessed information.
We demonstrate the ROB-contention attack on actual cores and show that, indeed, instructions after the reconvergence point of a control-speculation can leak information accessed during the control-speculation. To prevent ROB-contention attacks, assuming defenses such as DoM, we argue that frameworks that selectively lift such defenses must take into account Forward Speculative Interference and change their tactics. We propose a number of mitigations that we are in the process of evaluating.

\section*{Acknowledgments}

This work was supported by Microsoft Research through its EMEA PhD
Scholarship Programme grant no. 2021-020 and by the Swedish Research
Council grants no. 2015-05159 and 2018-05254.

\bibliographystyle{IEEEtranS}
\bibliography{defs.bib, refs.bib}

\begin{thebibliography}{10}
\providecommand{\url}[1]{#1}
\csname url@samestyle\endcsname
\providecommand{\newblock}{\relax}
\providecommand{\bibinfo}[2]{#2}
\providecommand{\BIBentrySTDinterwordspacing}{\spaceskip=0pt\relax}
\providecommand{\BIBentryALTinterwordstretchfactor}{4}
\providecommand{\BIBentryALTinterwordspacing}{\spaceskip=\fontdimen2\font plus
\BIBentryALTinterwordstretchfactor\fontdimen3\font minus
  \fontdimen4\font\relax}
\providecommand{\BIBforeignlanguage}[2]{{%
\expandafter\ifx\csname l@#1\endcsname\relax
\typeout{** WARNING: IEEEtranS.bst: No hyphenation pattern has been}%
\typeout{** loaded for the language `#1'. Using the pattern for}%
\typeout{** the default language instead.}%
\else
\language=\csname l@#1\endcsname
\fi
#2}}
\providecommand{\BIBdecl}{\relax}
\BIBdecl

\bibitem{aimoniotis2021reorder}
P.~Aimoniotis, C.~Sakalis, M.~Sj{\"a}lander, and S.~Kaxiras, ``Reorder buffer
  contention: A forward speculative interference attack for speculation
  invariant instructions,'' \emph{IEEE Computer Architecture Letters}, vol.~20,
  no.~2, pp. 162--165, 2021.

\bibitem{muontrap20}
\BIBentryALTinterwordspacing
S.~{Ainsworth} and T.~M. {Jones}, ``Muontrap: Preventing cross-domain
  spectre-like attacks by capturing speculative state,'' in \emph{Proceedings
  of the International Symposium on Computer Architecture}, 2020, pp. 132--144.
  [Online]. Available: \url{https://doi.org/10.1109/ISCA45697.2020.00022}
\BIBentrySTDinterwordspacing

\bibitem{angerfog21-instructiontables}
\BIBentryALTinterwordspacing
F.~Anger, ``Instruction tables,'' May 2021. [Online]. Available:
  \url{https://www.agner.org/optimize/instruction_tables.pdf}
\BIBentrySTDinterwordspacing

\bibitem{angerfog21}
\BIBentryALTinterwordspacing
------, ``The microarchitecture of {Intel}, {AMD}, and {VIA} {CPUs}: An
  optimization guide for assembly programmers and compiler makers,'' May 2021.
  [Online]. Available:
  \url{https://www.agner.org/optimize/microarchitecture.pdf}
\BIBentrySTDinterwordspacing

\bibitem{barber_specshield_2019}
\BIBentryALTinterwordspacing
K.~Barber, A.~Bacha, L.~Zhou, Y.~Zhang, and R.~Teodorescu, ``{SpecShield}:
  {Shielding} {Speculative} {Data} from {Microarchitectural} {Covert}
  {Channels},'' in \emph{Proceedings of the International Conference on
  Parallel Architectural and Compilation Techniques}, Sep. 2019, pp. 151--164.
  [Online]. Available: \url{https://doi.org/10.1109/PACT.2019.00020}
\BIBentrySTDinterwordspacing

\bibitem{specinterference21}
\BIBentryALTinterwordspacing
M.~Behnia, P.~Sahu, R.~Paccagnella, J.~Yu, Z.~N. Zhao, X.~Zou,
  T.~Unterluggauer, J.~Torrellas, C.~Rozas, A.~Morrison \emph{et~al.},
  ``Speculative interference attacks: Breaking invisible speculation schemes,''
  in \emph{Proceedings of the Architectural Support for Programming Languages
  and Operating Systems}, Apr. 2021, pp. 1046--1060. [Online]. Available:
  \url{https://doi.org/10.1145/3445814.3446708}
\BIBentrySTDinterwordspacing

\bibitem{safespec_2019}
K.~N. Khasawneh, E.~M. Koruyeh, C.~Song, D.~Evtyushkin, D.~Ponomarev, and
  N.~Abu-Ghazaleh, ``{SafeSpec}: {Banishing} the {Spectre} of a {Meltdown} with
  {Leakage}-{Free} {Speculation},'' in \emph{Proceedings of the {ACM}/{IEEE}
  Design Automation Conference}, Jun. 2019, pp. 1--6.

\bibitem{kocher_spectre_2018}
\BIBentryALTinterwordspacing
P.~Kocher, D.~Genkin, D.~Gruss, W.~Haas, M.~Hamburg, M.~Lipp, S.~Mangard,
  T.~Prescher, M.~Schwarz, and Y.~Yarom, ``Spectre attacks: Exploiting
  speculative execution,'' in \emph{Proceedings of the {IEEE} Symposium on
  Security and Privacy}, May 2019, pp. 19--37. [Online]. Available:
  \url{https://doi.org/10.1109/SP.2019.00002}
\BIBentrySTDinterwordspacing

\bibitem{Saileshwar19}
\BIBentryALTinterwordspacing
G.~Saileshwar and M.~K. Qureshi, ``{CleanupSpec}: An "undo" approach to safe
  speculation,'' in \emph{Proceedings of the {ACM/IEEE} International Symposium
  on Microarchitecture}, 2019, pp. 73--86. [Online]. Available:
  \url{http://doi.acm.org/10.1145/3352460.3358314}
\BIBentrySTDinterwordspacing

\bibitem{ghosts}
\BIBentryALTinterwordspacing
C.~Sakalis, M.~Alipour, A.~Ros, A.~Jimborean, S.~Kaxiras, and M.~Sj\"alander,
  ``Ghost loads: What is the cost of invisible speculation?'' in
  \emph{Proceedings of the {ACM} International Conference on Computing
  Frontiers}, 2019, pp. 153--163. [Online]. Available:
  \url{http://doi.acm.org/10.1145/3310273.3321558}
\BIBentrySTDinterwordspacing

\bibitem{dom19}
\BIBentryALTinterwordspacing
C.~Sakalis, S.~Kaxiras, A.~Ros, A.~Jimborean, and M.~Sj\"{a}lander, ``Efficient
  invisible speculative execution through selective delay and value
  prediction,'' in \emph{Proceedings of the International Symposium on Computer
  Architecture}, 2019, pp. 723--735. [Online]. Available:
  \url{http://doi.acm.org/10.1145/3307650.3322216}
\BIBentrySTDinterwordspacing

\bibitem{taram_context-sensitive_2019}
\BIBentryALTinterwordspacing
M.~Taram, A.~Venkat, and D.~Tullsen,
  ``\BIBforeignlanguage{en}{Context-{Sensitive} {Fencing}: {Securing}
  {Speculative} {Execution} via {Microcode} {Customization}},'' in
  \emph{\BIBforeignlanguage{en}{Proceedings of the Architectural Support for
  Programming Languages and Operating Systems}}, 2019, pp. 395--410. [Online].
  Available: \url{http://dl.acm.org/citation.cfm?doid=3297858.3304060}
\BIBentrySTDinterwordspacing

\bibitem{yan_invisispec:MICRO2018}
\BIBentryALTinterwordspacing
M.~Yan, J.~Choi, D.~Skarlatos, A.~Morrison, C.~W. Fletcher, and J.~Torrellas,
  ``{InvisiSpec}: Making speculative execution invisible in the cache
  hierarchy,'' in \emph{Proceedings of the {ACM/IEEE} International Symposium
  on Microarchitecture}, Oct. 2018, pp. 428--441. [Online]. Available:
  \url{https://doi.org/10.1109/MICRO.2018.00042}
\BIBentrySTDinterwordspacing

\bibitem{yarom_flush+_2014}
\BIBentryALTinterwordspacing
Y.~Yarom and K.~Falkner, ``{FLUSH}+ {RELOAD}: A high resolution, low noise, l3
  cache side-channel attack,'' in \emph{Proceedings of the {USENIX} Security
  Symposium}, 2014, pp. 719--732. [Online]. Available:
  \url{https://www.usenix.org/conference/usenixsecurity14/technical-sessions/presentation/yarom}
\BIBentrySTDinterwordspacing

\bibitem{sdo20}
J.~{Yu}, N.~{Mantri}, J.~{Torrellas}, A.~{Morrison}, and C.~W. {Fletcher},
  ``Speculative data-oblivious execution: Mobilizing safe prediction for safe
  and efficient speculative execution,'' in \emph{Proceedings of the
  International Symposium on Computer Architecture}, 2020, pp. 707--720.

\bibitem{yu_speculative_2019}
\BIBentryALTinterwordspacing
J.~Yu, M.~Yan, A.~Khyzha, A.~Morrison, J.~Torrellas, and C.~W. Fletcher,
  ``Speculative {Taint} {Tracking} ({STT}): {A} {Comprehensive} {Protection}
  for {Speculatively} {Accessed} {Data},'' in \emph{Proceedings of the
  {ACM/IEEE} International Symposium on Microarchitecture}, 2019, pp. 954--968.
  [Online]. Available: \url{http://doi.acm.org/10.1145/3352460.3358274}
\BIBentrySTDinterwordspacing

\bibitem{invarspec20}
\BIBentryALTinterwordspacing
Z.~N. Zhao, H.~Ji, M.~Yan, J.~Yu, C.~W. Fletcher, A.~Morrison, D.~Marinov, and
  J.~Torrellas, ``Speculation invariance ({InvarSpec}): Faster safe execution
  through program analysis,'' in \emph{Proceedings of the {ACM/IEEE}
  International Symposium on Microarchitecture}, 2020, pp. 1138--1152.
  [Online]. Available: \url{https://doi.org/10.1109/MICRO50266.2020.00094}
\BIBentrySTDinterwordspacing

\end{thebibliography}

\end{document}